\documentclass[referee]{raa}            

\usepackage{graphicx,times}
\usepackage{natbib}

\usepackage{soul}
\setstcolor{blue}

\providecommand{\bm}[1]{\mbox{\boldmath$#1$\unboldmath}}

\begin{document}

 \title{Magneto-acoustic oscillations observed in a solar plage region}
\volnopage{ {\bf 2012} Vol.\ {\bf 9} No. {\bf XX}, 000--000}
\setcounter{page}{1}

\author{Haisheng Ji\inst{1, 2}, Parida Hashim\inst{2, 3}, Zhenxiang Hong\inst{2},  Zhe, Xu\inst{2}, Jinhua, Shen\inst{3},  Kaifan Ji\inst{5}, Wenda, Cao\inst{6}}
\institute{University of the Chinese Academy of Sciences, Beijing, 100871, China \\
 \and  Purple Mountain Observatory, DMSA/CAS, Nanjing, China, correspondence: jihs@pmo.ac.cn \\
 \and  Xinjiang Astronomical Observatory, CAS, Urumqi, 830011, China\\
 \and  Yunnan Astronomical Observatory, CAS, Kunming, 650216, China  \\
 \and Big Bear Solar Observatory, 40386 North Shore Lane, Big Bear City, CA 922314, USA \\
            \vs \no
   {\small Received [2020] [xx] [xx]; accepted [xxxx] [xx] [xx]}
}

\abstract{We gave an extensive study for the quasi-periodic perturbations on the time profiles of the line of sight (LOS) magnetic field in $10 \times 10$ sub-areas in a solar plage region (corresponds to a facula on the photosphere). The perturbations are found to be associated with enhancement of He I 10830 \AA\ absorption in a moss region, which is connected to loops with million-degree plasma. FFT analysis to the perturbations gives a kind of spectrum similar to that of Doppler velocity: a number of discrete periods around 5 minutes. The amplitudes of the magnetic perturbations are found to be proportional to magnetic field strength over these sub-areas. In addition, magnetic perturbations lag behind a quarter of cycle in phase with respect to the p-mode Doppler velocity. We show that the relationships can be well explained with an MHD solution for the magneto-acoustic oscillations in high-$\beta$ plasma. Observational analysis also shows that, for the two regions with the stronger and weaker magnetic field, the perturbations are always anti-phased. All findings show that the magnetic perturbations are actually magneto-acoustic oscillations on the solar surface, the photosphere, powered by p-mode oscillations. The findings may provide a new diagnostic tool for exploring the relationship between magneto-acoustic oscillations and the heating of solar upper atmosphere, as well as their role in helioseismology. 
}
\authorrunning{Ji et al.}
\titlerunning{Magneto-acoustic oscillations}
\maketitle
\newpage

\section{Introduction}

The solar chromosphere is an interface layer, through which the mass and energy flows pass from below to enable plasma circulation \citep{2020RAA....20..166C} and to heat the upper atmosphere \citep{Aschwanden2007, DePontieu2009, iris}. An important target for studying the interface layer is the footpoint region of coronal loops, shown as plages in the chromosphere (or facula regions of the photosphere). The region is prevalent with oscillations with both 3 and 5 minutes periodicities \citep{2003ApJ...587..806M, DePontieu03, 2003ApJ...585.1138H, Wiehr1985, Tian2014Sci...346A.315T, 2001ApJ...554..424J}.  In a plage area, of particular interest is the so-called EUV ``moss'', a region being connected to coronal loops with million-degree hot plasma \citep{Berger1999}.  In this region, much stronger heating rate is believed to be constantly occurring.  Nevertheless, the region has its name since it is full of dark inclusions from low temperature plasma making it take the appearance of reticulated bright EUV emission \citep{FletcherDePontieu1999}.  The dark inclusions are found to jointly appear and disappear, a signature of oscillatory fine-scale mass and energy flows going upward \citep{FletcherDePontieu1999, DePontieu03, DePontieu2006}.  With high-resolution narrow-band imaging at He \textsc{i} 10830 \AA\ for an EUV moss inside a plage region \citep{jicaogoode2012}, Hashim et al. (2020) reported correlations between EUV emissions and  magnetic perturbations with the period of $\sim$ 5 minutes.  Understanding the nature of these magnetic perturbations and their coupling with underlying global p-mode is an important topic since they are related to heating of the upper atmosphere. 

Soon after Severny (1967) made the first attempt to identify magnetic perturbations as MHD waves in the solar atmosphere, Tanenbaum, Wilcox and Howard (1971) reported the existence of periodic oscillations of magnetic field related to p-mode. Since then, there have appeared many research works on magnetic field oscillations, mostly on the oscillations around the area of a pore or a sunspot (Khomenko \& Collados 2015; Bogdan \& Judge 2006; Staude 2002 and references therein). The measured periods for the oscillations of the magnetic field in sunspots are centered around 3 or 5 minutes and amplitudes range from a few Gauss in most cases up to tens of Gauss  by some authors (e.g., Horn et al. 1997; R{\"u}edi et al. 1998; Kupke et al. 2000; Bellot Rubio et al. 2000; Balthasar 1999; Zhugzhda et al. 1983). The low S/N ratio resulting from the small amplitude of magnetic field oscillations leads to contradictory results among different authors. With a set of well-observed sunspot data, Lites et al. (1998) gave an upper limit of about 4 G for the amplitude of the magnetic field oscillations, for which they regarded as of instrumental effects. Furthermore, some authors consider the measured fluctuations to be the results of cross-talk with velocity and intensity, including the opacity effect (R{\"u}edi et al. 1999, Bellot Rubio et al. 2000; R{\"u}edi \& Cally 2003; Khomenko et al. 2003; Zhao \& Chen 2018).  

 To identify true magnetic oscillations and exclude the possibility of cross-talk with the p-mode Doppler velocity, phase difference between them is an important parameter (e.g., Fujimura \& Tsuneta 2009). For magneto-acoustic oscillations around solar disk center, Urich (1996) anticipated and observed a $\sim \pi/2$ phase difference between the p-mode upward Doppler velocity and magnetic field variations in the photosphere. The $\sim \pi/2$ phase difference for sausage-mode waves was worked out in the MHD framework and observed in well-observed pores by Freij et al. (2016).  It was also observed in many other investigations \citep{Norton1999, Ruedi1998, BellotRubio2000, Fujimura2009}. However, R{\"u}edi et al. (2000) and Norton et al. (2001) obtained the phase difference of $-\pi/2$. In addition, some authors still suggested that the observed $ \pi/2$  phase difference is due to the opacity fluctuations caused by p-mode velocity field \citep{Lites1998, BellotRubio2000, Ruedi1998}.  Fujimura \& Tsuneta (2009) explained the $ \pi/2$ phase-difference as the result of superposition of the ascending wave and the descending wave reflected at chromosphere/corona boundary. 
 
 Beside, there is a phenomenon of acoustic absorption by sunspots, ``p-mode absorption'' , as reported by Braun, Duvall and LaBonte (1988). Theoretical models usually explained it as the result of conversion of fast-mode to slow-mode by the oscillations of vertical magnetic fields within sunspots \citep{KhomenkoandCollados2015, 1992ApJ...391L.109S, 2006RSPTA.364..447R, 1988ApJ...335.1015B, 1993ApJ...402..721C}.  In these models, the trapped fast-mode waves experience reflections at the ends of the vertical magnetic fields due to the increase of Alfv\'{e}n speed with height and the increase of the acoustic speed with depth (Khomenko \& Collados 2015). The trapped waves become partly absorbed on their passages around the layer of $\beta \approx 1$.  In the theoretical models given by Cally and Bogdan (1993) and Cally et al. (1994), the appearance of complex frequencies or wave numbers successfully predict the absorption. A similar but more refined model given by Spruit and Bogdan (1992) predicted some characteristic signature of absorption for the f-mode and along p-mode ridge and provided the diagnostic possibility to determine the sunspot magnetic field strength from the location in wavenumber of the predicted absorption minima. For a coherent observational picture and related models (theories, as well as numerical simulations) for the oscillations in sunspots, readers can refer to some review papers (e.g., Khomenko \& Collados 2015; Jess et al. 2015, Bogdan \& Judge 2006). Note that nearly all the models were given to account for the magnetic oscillations in sunspots at a layer of which thermal pressure and magnetic pressure are balanced (i.e. plasma $\beta \approx 1$). 
 
In his paper, we give a detailed observational analysis to magnetic oscillations in a plage region, the same plage analyzed by Hashim et al. (2020). We take a qualitative approach to understand the observed phenomena in an MHD framework, assuming that plasma $\beta$ is much larger than 1. In \S \ 2, we give a solution for magneto-acoustic waves for high plasma $\beta$ regions filled by a vertical magnetic field with horizontal gradient. In \S \ 3,  after a brief introduction to the the quasi-periodic He \textsc{i} 10830 \AA\ absorption in a moss region, we give a detailed analysis for the oscillations of the line of sight (LOS) magnetic field and Doppler velocity, and their correlations or coupling.  Conclusions and discussions are given in \S \ 4.

 \section{Magneto-acoustic oscillations in high-$\beta$ plasma inside a static vertical magnetic field with horizontal gradient}

In most area of the photosphere,  the $\beta$ value of the plasma, the ratio of thermal pressure over magnetic pressure, is much larger than 1 (see \S \ 3 for an estimate). Studies of magneto-acoustic waves in flux tubes in high-$\beta$ photosphere have been carried out by many authors (e.g., Spruit 1982; Edwin \& Roberts 1983; Ulmschneider et al. 1991). In this paper, we will have a different approach to take the advantage of the high-$\beta$ nature of the photosphere. To model the line of sight magnetic field $B_{LOS}$ in a facula regions of the photosphere, we take a cylindrical coordinate system and assume a vertical magnetic field $B_0(r) \hat{z} $ with radial (horizontal) inhomogeneity. We neglect any azimuthal variations, which is equivalent to the sausage mode with $m = 0$. 

We start with two ideal MHD equations (momentum and induction) neglecting gravity, viscosity, and diffusion. The momentum and induction equations are given by 
\begin{equation}
 \rho \left[\frac{\partial}{\partial t} + (\vec{v}\cdot\vec{\nabla})\right]\vec{v} = - \vec{\nabla}P + \frac{1}{\mu_0} (\vec{\nabla}\times\vec{B})\times\vec{B}, 
\label{QJ001}
\end{equation}
 \begin{equation}
\frac{\partial \vec{B}}{\partial t} = \vec{\nabla}\times(\vec{v}\times\vec{B}).
\label{QJ002}
\end{equation}
We then introduce linear perturbations so that $\vec{v} = \vec{v}_1, \vec{B} = B_0\hat{z} + \vec{B}_1, P = P_0 + P_1$, where 
the subscript ``1" indicates perturbed properties. By assuming a static background plasma, the linearized induction equation and momentum equation can be written as:
\begin{equation}
\frac{\partial \vec{B}_1}{\partial t} = \frac{\partial \vec{v}_1}{\partial z} B_0 - 
(\vec{\nabla}\cdot\vec{v}_1)B_0\hat{z} - (\vec{v}_{1r} \frac{\partial }{\partial r})B_0\hat{z},
\label{QJ003}
\end{equation}
\begin{equation}
\rho_0 \frac{\partial \vec{v}_1}{\partial t}  = - \vec{\nabla} \left(P_1 + \frac{1}{\mu_0} B_0B_{1z}\right) +  \frac{\partial \vec{B}_1}{\partial z}\frac{B_0}{\mu_0} + \frac{1}{\mu_0} (\vec{B}_{1r} \frac{\partial }{\partial r}) B_0\hat{z},
\label{QJ004}
\end{equation}
We can assume wave solutions to Equations (\ref{QJ003} - \ref{QJ004}) to be in the following form:
\begin{equation}
\left\{
\begin{array}{ll}
\vec{v}_1 = & [v_{1r}(r) \hat{r} + v_{1z}(r) \hat{z}] e^{i(\omega t - k_z z)} \\
\vec{B}_1 = & [B_{1r}(r) \hat{r} + B_{1z}(r) \hat{z}] e^{i(\omega t - k_z z)} \\
P_1 = & p_1(r) e^{i(\omega t - k_z z)}
 \end{array}
 \label{xxq}
 \right .
\end{equation}
where $k_z$ is the wave number of the perturbation in the vertical direction, and $\omega$ is the angular frequency of the wave. In terms of the horizontal gradient of the vertical magnetic field, we have neglected horizontal components of the wave number with the assumption that $|\frac{\partial }{\partial r}| \gg k_{\perp}$. 

Since the horizontal components of Equation (\ref{QJ003}) gives that $\partial \vec{B}_{1}/\partial t = (\partial \vec{v}_{1}/\partial z) B_0$, the perturbed momentum equations for the horizontal components can be re-written as
\begin{equation}
\frac{\partial B_{1z}}{\partial r}  = -\frac{\mu_0}{B_0}\left[i \omega \rho_0 (1 - \frac{v_a^2}{v_p^2}) \vec{v}_{1r} + \frac{\partial p_1}{\partial r} \right] - B_{1z} \frac{\partial {\rm ln}B_0}{\partial r},
\label{QJ009}
\end{equation}
where $v_a \equiv \sqrt{B_0^2/(\mu_0\rho_0)}$ is the Alfv\'{e}n speed, 
and $v_p \equiv \omega/k_z$ is the phase speed of the magneto-acoustic wave in $z$-direction. 

We have assumed that the plasma $\beta \gg 1$ for facula regions, the fast-mode wave along the equilibrium field $B_0\hat{z}$ 
is the sound wave, and its phase speed $v_p^2 = c_s^2 = \gamma p_0/\rho_0$, 
where $\gamma$ is the ratio of the specific heats. In this case, the term 
$v_a^2/v_p^2 = 1/(2\gamma \beta) \ll 1$. Therefore, we are left with
\begin{equation}
\frac{\partial B_{1z} }{\partial r}  = -\frac{\mu_0}{B_0}\left(i \omega \rho_0 \vec{v}_{1r} + \frac{\partial p_1}{\partial r} \right) - B_{1z} \frac{\partial {\rm ln}B_0}{\partial r}.
\label{QJ010A}
\end{equation}
We can re-write the above equation as:
\begin{equation}
\frac{\partial B_{1z} }{\partial r}   = -\frac{\mu_0}{B_0}(\rho_0 \frac{\partial \vec{v}_{1}}{\partial t} + \vec{\nabla}p_1 )_{r} - B_{1z} \frac{\partial {\rm ln}B_0}{\partial r}.
\label{QJ12}
\end{equation}
The above relation may be used to evaluate magnetic perturbations with transverse gradient for static vertical magnetic field being included. In the region with high plasma $\beta$-value,  gas pressure dominates the magnetic pressure. In this case, we have
\begin{equation}
\lim_{B_0 \rightarrow 0} (\rho_0 \frac{\partial \vec{v}_{1}}{\partial t} + \vec{\nabla}p_1)  = 0 
\label{QJ12cJ}
\end{equation}
For the sake of simplicity, we can define a function as following
\begin{equation}
 W(r) = (\rho_0 \frac{\partial \vec{v}_{1}}{\partial t} + \vec{\nabla}p_1 )_{r}.
\label{weak}
\end{equation}
We see that $W(r)$ is a kind of function that can be used to describe the degree of deviation of magnetized high-$\beta$ plasma from non-magnetized one (pure gas).  For pure gas $W(r) = 0$, which means that perturbing thermal pressure gradient is balanced by the change of perturbed Doppler velocity of the gas. The presence of magnetic field in a plasma will produce a surplus value for W(r). Therefore, in high-$\beta$ plasma, W(x) is a kind of source function that holds a positive correlation with the strength of magnetic field. Then, the amplitude of the perturbed magnetic field in z-direction can be given in the following way
$$ B_{1z}= \left[C - \mu_0 B_0^{-1} \int Wdr \right],$$
where $C$ is an integral constant. Since the perturbed magnetic field $B_{1z}$ vanishes when the source term $W(x)$ approaches zero, i.e., $B_{1z} = 0$ when $W = 0$,  the integral constant becomes zero. In the end, the perturbed magnetic field in z-direction is given by
 \begin{equation}
B_{1z} = -\mu_0 B_{0}^{-1}  \int W d r.
 \label{v12T}
\end{equation}
 Equation (\ref{v12T}) gives a solution that can be use to evaluate the perturbation of LOS magnetic field in high-$\beta$ plasma. Though $B_0$ term appears in the denominator, the oscillation amplitude indeed vanishes when $B_0 = 0$.  In this sense, solution (\ref{v12T}) shows that, magnetic oscillation amplitude becomes larger when magnetic field becomes slightly stronger.  We will discuss this in \S ~4.

\begin{figure}[h]
\includegraphics[width=9cm]{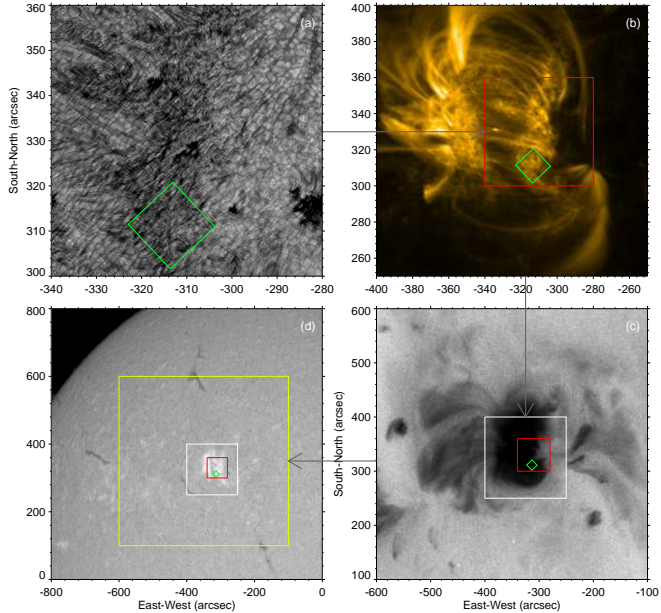}
\caption{\small Panels a-d show the plage region observed at He \textsc{i} 10830 \AA, Fe IV 171 \AA, Soft X-ray, and H$_\alpha$ 6563 \AA, respectively.  The panels are arranged in such a way that the field of view is progressively zooming in to give the position of the moss region (inside the green-colored box) in the plage, and the position of the plage on solar disk. The area for investigating magneto-acoustic oscillations (see Fig.  2) is inside the white boxes, which is also the field of view of panel b.} 
\label{overview}
\end{figure}

\section{Observation and results}

\begin{figure}
\includegraphics[width=9cm]{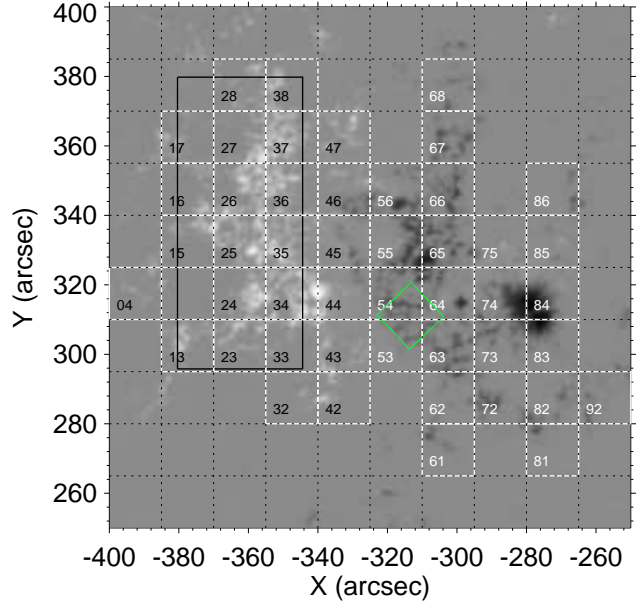}
\caption{The magnetogram and the $15" \times 15"$ sub-areas used to obtain time profiles for mean magnetic field and Doppler velocity. The field of view is the same as the area inside the white boxes in Fig.  1. The area inside the black box is given an additional analysis by isolating the stronger magnetic pixels. The result of the additional analysis is given in Fig. 10.}
\label{magnetogram}
\end{figure}

For completeness of the paper, we give here the necessary context of observations for the following analysis of magnetic field oscillations. A more extended analysis for the plage region has been given by Hashim et al. (2020). Fig. \ref{overview} gives an overview of the plage region observed with different telescopes. The field of view of high-resolution observations at He \textsc{i} 1083 nm covers one footpoint region of a coronal arcade in the active region NOAA 11259 (Fig.  \ref{overview}a-b) as observed by the Atmospheric Imager Assembly (AIA) on board the Solar Dynamic Observatory (SDO) \citep{aia-sdo, SDO}. Fig.  \ref{overview}c gives its appearance taken with the Ti-poly filter by the X-ray telescope (XRT) on board Hinode \citep{Kosugi2007, Hinode}, showing that the arcade contains plasma of a few million degrees.  In an H$_\alpha$ image, the footpoint regions are shown as plages near the active region NOAA 11259 (Fig. \ref{overview}d). The plages are of opposite magnetic polarities when being compared with a  corresponding LOS magnetogram (See Fig.  \ref{magnetogram}). The magnetograms, as well as the Dopplergrams analyzed in this paper, are obtained from the observations made by the Helioseismic Magnetic Imager (HMI) \citep{hmi-sdo} on-board SDO. HMI observes the full disk Sun in the Fe I absorption line at 6173 \AA\ to measure oscillations of Doppler velocity and the magnetic field in the photosphere. It provides full-disk, high-cadence Doppler, intensity, and magnetic images at 1 arcsec resolution (4096$\times$4096-pixel images) of the solar photosphere. The time period of downloaded magnetograms is from 17:40 to 22:00 UT and the time cadence is 45 seconds. 

\begin{figure}
\includegraphics[width=11cm]{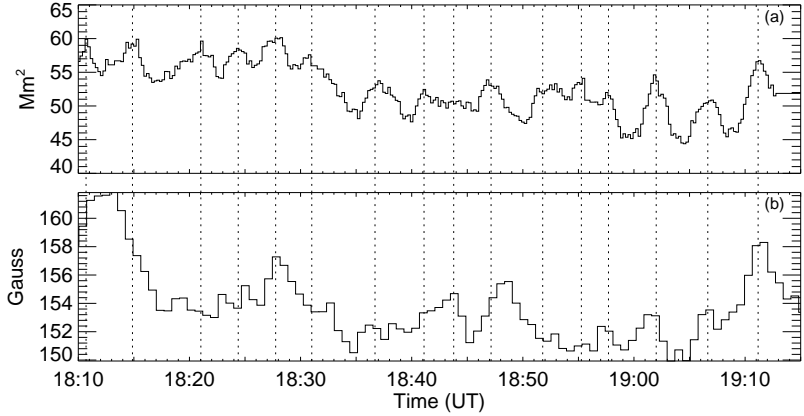}
\caption{\small The upper panel gives the time profile of the total area obtained by counting the pixels with enhanced He \textsc{i} 10830 \AA\ absorption in the moss region. Panel (b) is the time profile for the mean magnetic field in the same region.}
\label{mossabsorption}
\end{figure}

The initial area of interest is an EUV moss region with the size of $\sim$ $10 \times 10$ Mm$^2$ (inside the green boxes of Fig.  \ref{overview}).  With He \textsc{i} 10830 \AA\ filtergrams,  we divide the moss area into two distinct regions \citep{Hong2017}: patches with enhanced He \textsc{i} 10830 \AA\ absorption with emission less than $4.5 \times 10^3$ counts per pixel (EAPs: enhanced absorption patches) and patches with less absorption. By totaling all pixels in EAPs, we obtain the time profile for the total area of He \textsc{i} 10830 \AA\ absorption in the moss region, and its variation is given in Fig.  \ref{mossabsorption}a. We see that the absorption in the moss region shows a periodic oscillating nature. The result basically agrees with many previous results for solar EUV moss regions (e.g., De Moortel \& Nakariakov 2012). The peaks for the He \textsc{i} 10830 \AA\ absorption actually represent periodic tiny heating events in the moss region \citep{Hong2017}. To explore their association with perturbations of magnetic field, we compare it with the time profile of mean LOS magnetic field (being equivalent to net magnetic flux) in the same region (Fig.  \ref{mossabsorption}b). We see that some absorption peaks are obviously coincident with the peaks, though being weak, on the time profile for the magnetic field. The coincidence strongly suggests the existence of magnetic oscillations in the plage region and, also, its importance for solving the problem of coronal heating. 

In order to investigate the magnetic perturbations, we also select other regions, which include the whole plage regions, the sunspot and surrounding quiet regions with the weakest magnetic field (Fig.  \ref{magnetogram}). To investigate general nature for the perturbations of magnetic field, we divide the whole area into $10 \times 10$  sub-areas. Then, with co-aligned magnetograms, we get the time profile of mean magnetic field (equivalent to net magnetic flux) in each sub-area.  Fig.  \ref{magnetogram} shows the uniformly divided sub-areas, with the numbered ones highlighting the stronger magnetic field in the plage and sunspot regions.    

\begin{figure}
\includegraphics[width=11cm]{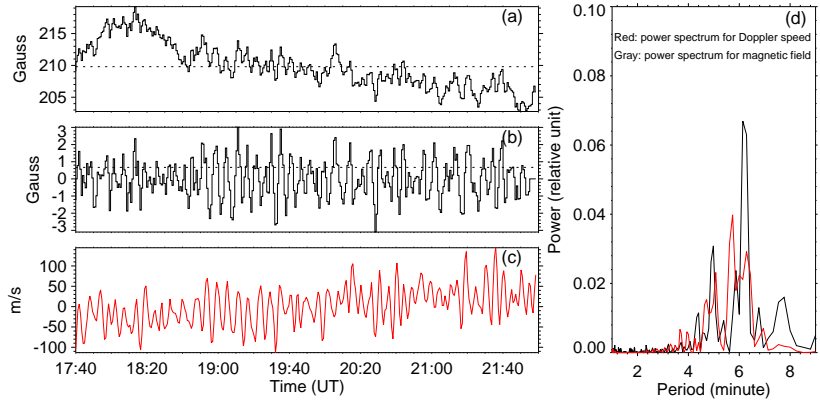}
\caption{An example showing magneto-acoustic oscillations in sub-area 37 in Fig.  2. a) The time profile of the mean magnetic field in the sub-area. b) A train of oscillatory components obtained from the time profile in panel (a) by subtracting its slowly varying component (smoothed one). c)  Time profile of Doppler velocity (blue shifted) in the same sub-area. d) The power spectra, in relative unit, for the oscillatory components in panels b and c. On-line animation of this figure is available.}
\label{subarea37}
\end{figure}

As a demonstration, Fig.  \ref{subarea37} gives sample results from sub-area 37.  The mean time profile for the mean magnetic field and its fast-varying components are given in panels a-b. The fast-varying components were obtained by subtracting the slowly-varying component, the smoothed one by 11-point running averaging to the original time profile. We see that quasi-periodic magnetic perturbations persistently appear for the sub-area. To explore the relationship with the global p-mode oscillations, Fig. \ref{subarea37}c gives the time profile of the mean blue shifted Doppler velocity in the same sub-area. The Dopplergrams used here are calibrated with most of the observer motion effects, solar-rotation signal and background being carefully removed. We see that the magnetic oscillations seem to be synchronized with the oscillations of the Doppler velocity.   

\begin{figure}   
\includegraphics[width=11cm]{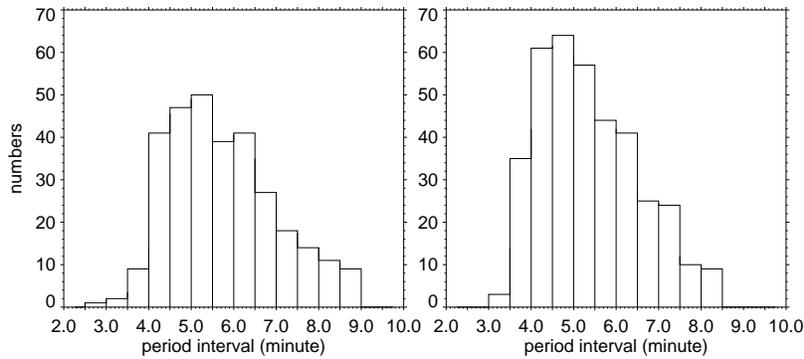}
\caption{Histograms showing the distribution of periods of magnetic field perturbations (left panel) and Doppler velocity (right panel).  All periods are obtained from the power spectra for all numbered sub-area in Fig.  2.}
\label{histogram}
\end{figure}

We carried out Fast Fourier Transforming to both kinds of time profiles in the numbered sub-areas. The right panel of Fig. \ref{subarea37} gives two sample power spectra (in relative units) for sub-area 37. The distributions of the two kinds of spectra are similar, showing a series of discrete periods. However, the periods for  the perturbations of magnetic field and Doppler velocity are not coincide completely. For all numbered sub-areas, we get totally 310 and 373 periods from the power spectra of magnetic field and Doppler velocity respectively. The periods are obtained with visual inspection to the peaks which are above 95\% confidence level. Fig. \ref{histogram} gives two histograms for the distributions of the periods. We see that the two histograms are quite similar, with a maximum around the period of 5 minutes. It shows that the magnetic perturbations are intrinsically linked to the p-mode. It is also worth mentioning that magnetic field perturbations contain more components with longer periods.   

\begin{figure}
\includegraphics[width=11cm]{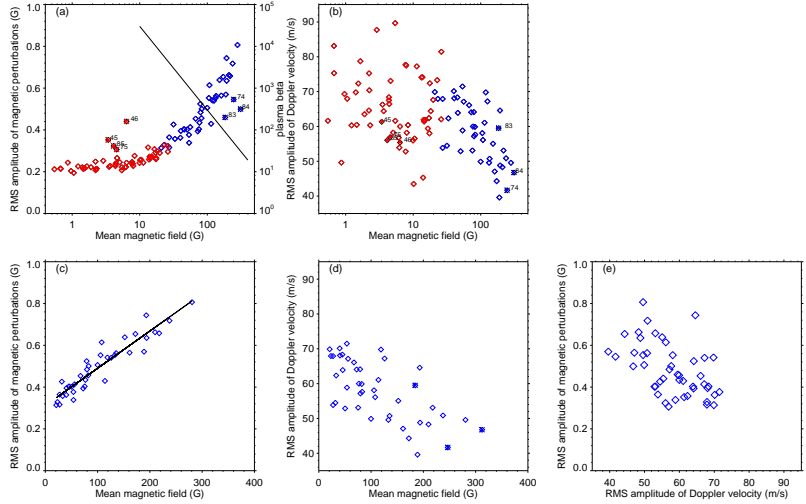}
\caption{Scatter plots (diamond symbols) show a number of mutual relationships among the RMS amplitudes of magnetic perturbations, the RMS amplitudes of Doppler velocity and the mean magnetic field strength with the results obtained from all sub-areas in Fig.  2.  Red diamond symbols represent the weak field sub-areas (not numbered), while blue diamond symbols are for numbered sub-areas. The over-plotted cross signs pick out the sub-areas with mixed polarities inside them (red cross) and the sub-areas overlapped with the sunspot (blue). The numbers beside the over-plotted cross signs give the corresponding sub-areas in Fig.  2.  In panel (c), we give a linear fit for all blue diamond symbols in panel (a) except the three with crossed symbols. Panel (d)  shows that Doppler velocity becomes smaller in the regions with the stronger magnetic field strength. Panel (e) gives the relationship between the RMS amplitudes of magnetic perturbations and Doppler velocity. The line in panel (a) gives the estimated plasma-$\beta$ values for the plage region with different magnetic field strength.}
\label{linear}
\end{figure}

For magnetic field perturbations, we measure the root-mean-square (RMS) amplitudes and magnetic field strength in each sub-area of Fig. \ref{magnetogram}.  Taking the time profile in Fig. \ref{subarea37}a as an example, the RMS amplitude and mean magnetic field strength (see the two dotted horizontal lines in Fig. \ref{subarea37}a-b) are $\sim$ 0.67 G and $\sim$ 209.8 G, respectively. A scatter plot for the amplitude and mean strength from all sub-areas is shown in Fig. \ref{linear}a, in which blue signs stand for all numbered sub-areas, while red signs are for the other sub-areas. We see that mean magnetic field strength varies up to two orders of magnitude in the whole area. The plot shows that the perturbation amplitude grows with the magnetic field. This actually occurs after the mean magnetic field is larger than $\sim$ 10 G (Fig. \ref{linear}a). Below 10 G, the nearly constant amplitude of $\sim$ 0.25 G should belong to a noise level. Sub-areas 45, 46, 75 and 85 (with cross signs) are the regions with mixed magnetic polarities, which will result in a decrease in net magnetic flux (or the mean field value). Sub-areas 74, 83, and 84 (also with cross signs) actually belong to the sunspot. In the sunspot region, plasma $\beta$ will become much lower due to the lower thermal pressure as well as the stronger magnetic field.  

Here, we estimate the $\beta$ values of the plasma in the facula regions. By taking the density and temperature in the faculae as their typical values at the optical depth $\tau \sim 1$ ($n_p \sim 1.0 \times 10^{22}$ $m^{-3}$ and $T \sim 5800 $ K), the result of plasma-$\beta$ values versus magnetic fields is over-plotted in Fig.  \ref{linear}a. We see that, in the region where the magnetic field is less than $\sim$ 400 G, the $\beta$ value can be regarded as much larger than 1.  

After removing the blue-colored points with cross signs as well as red-colored points, we can get a rough linear relationship for the oscillating amplitude and the mean magnetic field  (Fig \ref{linear}c).  The linear regression coefficient is obtained as 0.0018. Fig. \ref{linear}b gives a scatter plot for the relationship between mean magnetic field strength and the RMS amplitude of Doppler velocity obtained from all sub-areas. After we remove the points (red diamond signs) of much less magnetized sub-areas, the result (Fig. \ref{linear}d) shows that Doppler velocity becomes smaller in the regions with the stronger magnetic field strength. Combination of the results from Fig.  6c-d gives that the oscillation amplitude of magnetic field and the magnitude of Doppler velocity is roughly anti-correlated over these sub-areas.  Fig.  6e shows the anti-correlation relationship. 

\begin{figure}
\includegraphics[width=11cm]{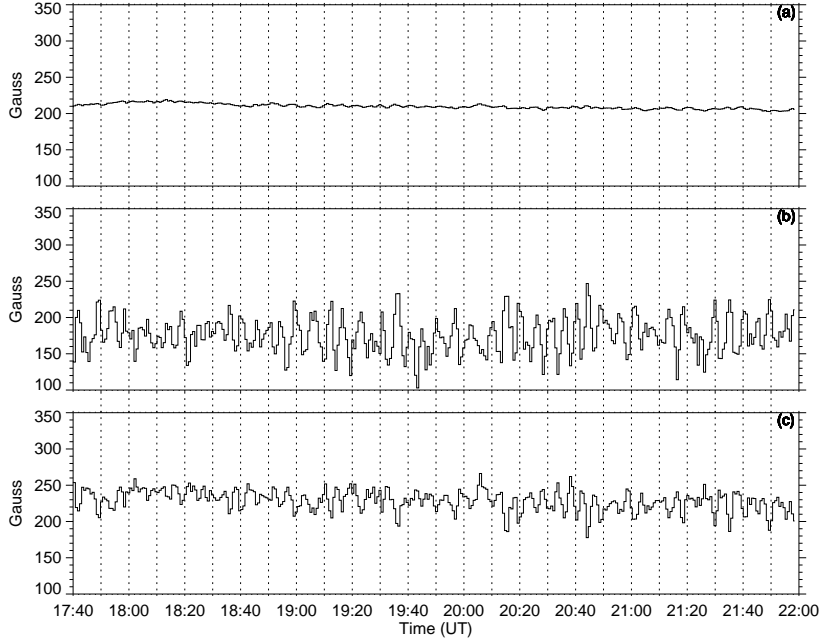}
\caption{Upper panel: the time profile of the mean magnetic field in sub-area 37 (the same as in Fig.  4a but with the larger value range for the vertical axis). Panel (b) gives the time profile for magnetic field averaged over pixels with Doppler velocity larger than 250 m s$^{-1}$, while panel (c) gives the time profile for magnetic field averaged over pixels with Doppler velocity slower than 250 m s$^{-1}$ in the same sub-area. Vertical lines is plotted to help readers to see the persistent anti-correlated phase relationship of the two kinds of time profiles. On-line animation for panels b-c is available showing the results from other labeled areas in Fig. 2.}
\label{antiphase-1}
\end{figure}

We further obtain two kinds of time profiles for the mean magnetic field strength, one is for those pixels with the larger Doppler velocity while another is for the smaller Doppler velocity. We find that their perturbations are always anti-phased. Taking sub-area 37 as an example for demonstration. The result is given in Fig. \ref{antiphase-1}b-c, in which the time profiles in panel b and panel c are associated with Doppler velocity amplitude larger than 250 m \ s$^{-1}$ and less than 250 m \ s$^{-1}$, respectively. Here, the selected value of 250 m \ s$^{-1}$ is not necessarily an accurate value. The two time profiles are constantly anti-phased. In Fig. \ref{antiphase-1}a, we redraw the time profile of mean magnetic field in sub-area 37 that is already given in \ref{subarea37}a, but with a much wider range of value for the vertical axis. The value range is the same as that of panels b-c for the purpose of comparison. In this way, the time profile in Fig. \ref{antiphase-1}a looks much more gradual. However, it splits into two anti-phased oscillating components with the larger amplitudes (Fig. \ref{antiphase-1}b-c) when we make a difference according to faster or slower Doppler velocity. 

\begin{figure}
\includegraphics[width=11cm]{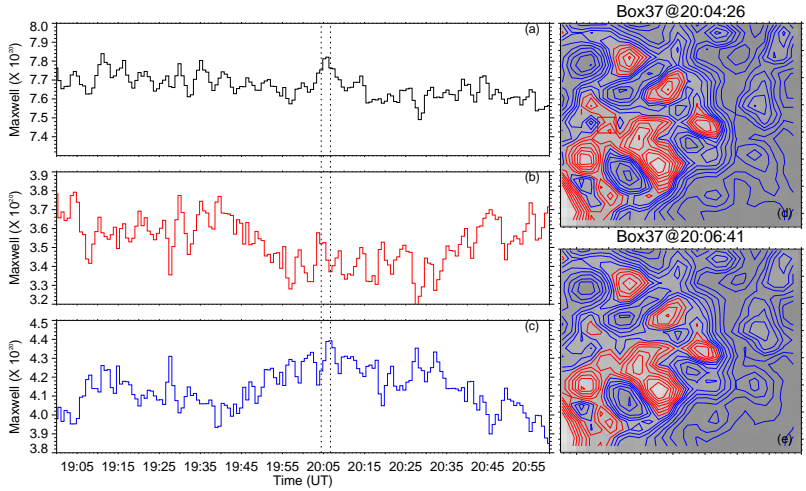}
\caption{Panel (a) is the time profile for total magnetic flux in sub-area 37, while panels (b) and (c) give the time profiles for magnetic flux over the pixels with magnetic field larger than 350 G and less than 350 G in the same sub-area. The right two panels show the spatial distribution (maps) of magnetic field at 20:04:26 and 20:06:41 UT. Red contour levels (370, 410, 450, 500, 550, 600, and 650 G) depict the spatial magnetic concentrations for the stronger magnetic field and blue contour levels (330, 290, 250, 200, 150, 110, 80, and 50 G) give progressively decreasing magnetic field toward their center. The two vertical dotted lines over the left panels correspond to the two timings of the right maps, representing alternating peak and valley times of strong magnetic flux and weak magnetic flux.  The red box in panel (d) in the small area of $3 \times 3$ pixels for a phase difference analysis, and the result is given in Fig.  9a. On-line animation of this figure is available.}
\label{antiphase-2}
\end{figure}

To scrutinize how one small region contains anti-phase perturbations of magnetic field, we divide sub-area 37 into two regions according to the magnitude of magnetic field. In this case, we obtain two kinds of time profiles for the magnetic flux, depending on wether the field strength is larger or less than a certain value, e.g., 350 G in this paper. The separation again divide the total magnetic flux in the sub-area into two anti-phased branches (Fig. \ref{antiphase-2}a-c). The right two panels of Fig. \ref{antiphase-2} show the spatial distribution for magnetic field.  They are at two adjacent peak and valley times of magnetic fluxes for the stronger and weaker magnetic field. From the contours, we see that, around the peak time of the stronger magnetic flux, the strengthening in magnetic concentration areas (with ever strengthening magnetic field toward the center of red contours) is actually accompanied by the weakening of the magnetically depressed area (with ever weakening magnetic field toward the center of blue contours). The enhancement in magnetic concentration areas and the weakening of the magnetically depressed area, which which are concurrent, are followed subsequent in-phase weakening and strengthening in the same locations.  The picture can be seen with the on-line animation more easily.  

\begin{figure}
\includegraphics[width=11cm]{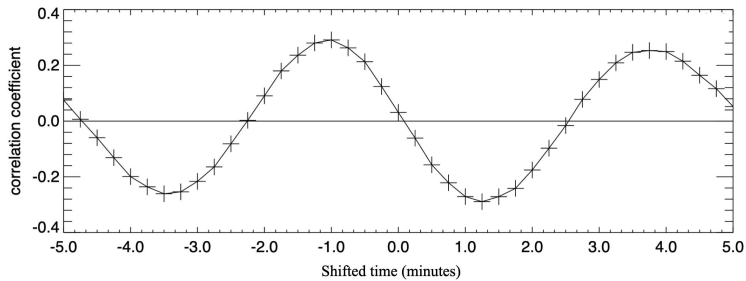}
\caption{The result of running correlation for the time profiles of velocity and magnetic field from the point with the small area of $3 \times 3$ pixels (see Fig.  8d) in sub-area 37.}
\label{cross}
\end{figure}

We have seen that the magnetic perturbations on the photosphere have the nature of magneto-acoustic waves being coupled to p-mode. For their phase relationship, we can understand why previous results in literature have been inconclusive, this may be just due to different observing apertures with insufficient spatial resolution. We see that, even within a small area, there exists anti-phased perturbations. Therefore phase comparison can only made in a much smaller area. We further divide sub-area 37 (totally $26 \times 26$ pixels in the sub-area) into $6 \times 6$ point-areas and carried out running-correlation analysis between the two kinds of perturbations in each area by shifting the time profile of magnetic oscillations back and forth. The time profiles are obtained in an area of $3 \times 3$ pixels$^2$ and the shifting is made with the accuracy of 15 seconds. Except for 16 point-areas mainly located on the right side of Fig. \ref{antiphase-2}d with the weaker magnetic field, the running-correlation gives a time difference of $\sim$ -1.2 minutes (Fig.  \ref{cross}) for maximum correlation. The 1.2-minute time difference corresponds the $\pi/2$ phase difference if we take the mean perturbation period as being 5 minutes.  During the cross-correlation analysis, Doppler velocity is multiplied by a factor -1, thus making upward Doppler velocity positive. Conventionally, blue-shifted (upward) Doppler velocity is recorded as negative, this will cause some confusion in phase analysis. For the above-mentioned $\pi/2$ phase difference between the perturbations of Doppler velocity and magnetic field, we would like to express it as $\phi_{-v_z} - \phi_{\delta B_{1z}} = +\pi/2$, with the plus or the minus sign being specially added to avoid any possible confusions. 

\section{Discussions and conclusions}

In this paper, we show that the magnetic perturbations found in the plage region are magneto-acoustic oscillations, cousins of p-mode oscillations.  The picture is supported from multiple aspects. At First, FFT analysis gives that the quasi-periodic perturbations of magnetic field contain a number of discrete periods that are quite similar to global p-mode oscillations. Secondly, we can use phase relationship to verify the picture.  For phase-difference, let us go back to previous section and rewrite equation (\ref{v12T}) but in the following form
 \begin{equation}
B_{1z} = \mu_0 B_{0}^{-1}  \int (e^{-i\frac{\pi}{2}} \omega \rho_0 \vec{v}_{1r} - \frac{\partial p_1}{\partial r}) dr.
 \label{v12Z}
\end{equation} 
In above equation, the terms $- \frac{\partial p_1}{\partial r}$ and $e^{-i\frac{\pi}{2}} \omega \rho_0 \vec{v}_{1r}$ are in phase due to the nature of high-$\beta$ plasma. Also, in high-$\beta$ cold plasma, we can assume that the vertical velocity is approximately in phase with the horizontal ones. In this way, equation (\ref{v12Z}) shows that there will be a $\pi/2$ phase difference between upward Doppler velocity and the observed perturbations of the LOS magnetic field. In other words, upward Doppler velocity will reach its maximum a quarter of a cycle before magnetic field does. This is supported by the observations given above. However, we have seen that the theoretical analysis is based on the assumption that the horizontal velocity and the vertical velocity are in phase. The validity requires further verification.

Also, we have seen that the $\pi/2$ phase-difference (as well as  the amplitude of magnetic oscillation) during data analysis depends on the size of the area of interest. The magneto-acoustic oscillations in the sub-areas of Fig. 2 can split into two anti-phased components if we divide them into two kind regions with the stronger magnetic field (the slower Doppler velocity) and the weaker magnetic field (the faster Doppler velocity) respectively. The two kinds of regions take their turn to have magnetic strengthening and weakening (Fig. 8). The synchronization between V and B in Fig. 4 is purely due to the fact that the size of each sub-area is still too large. From the results of Figs. 8 and 9, we may conclude that the critical resolution to get the right phase-difference in the plage region should be no larger than 2 arc sec.  The picture given in Fig. 8 may give the picture of sausage-mode slow waves, similar to the observational finding made by Freij et al. (2016). In addition, with data analysis to two well-observed pores, Freij et al. (2016) reported the $\pi/2$ phase difference and they reproduced the phase difference with an MHD model for sausage-mode. 

It is worth mentioning that the $\pi/2$ phase difference was predicted by Ulrich (1996) in a different way, his result was that Doppler velocity plays a $\pi/2$ phase-leading role. The phase difference has been observed in a number of papers as we have introduced in \S ~1  (e.g, Fujimura and Tsuneta 2009 and references therein), the results are varied. Note that most investigations in literature were carried out for sunspots where magnetic field plays a dominant role. It has been proposed that $\pi/2$ phase difference  could be caused by opacity fluctuations that move upward and downward the region where the spectral lines are sensitive to magnetic fields (e.g., Bellot Rubio et al. 2000).  However, Fujimura and Tsuneta (2009) ruled out the possibility of the opacity effect, and they propose that their observed phase difference ($\pi/2$) is consistent with the phase relation of the superposition of the ascending and descending kink waves. Further detailed analysis with well observed data is needed for this kind of research.

Third, we show that magnetic oscillation amplitude is larger in the regions with the stronger magnetic field, where the magnetized plasma has the larger deviation from pure gas. For a high-$\beta$ plasma fluid, the amplitude of magnetic field oscillations will decrease to zero when the magnetic field decreases to zero. In this way, Equation (9) allows us to use L'Hospital's rule to deal with  Equation (11), so we get 
$$ \frac{\partial } {\partial B_0} (\rho_0 \frac{\partial \vec{v}_{1r}}{\partial t} + \frac{\partial p_1}{\partial r} )|_{B_0 = 0}  = 0 $$
Combining with dimensional analysis, we may have
\begin{equation}
\rho_0 \frac{\partial \vec{v}_{1r}}{\partial t} + \frac{\partial p_1}{\partial r} = - \frac{\partial }{\partial r} (\sigma B^2_0),
 \label{deviation1}
\end{equation} 
in the vicinity of $B_0 = 0$, and $\sigma$ is a dimensionless parameter whose magnitude is much smaller than 1.  Substitute equation (\ref{deviation1}) into equation (\ref{v12T}), we have
\begin{equation}
B_{1z} = \sigma B_0.
\label{dBproptoB}
\end{equation} 
A statistics from a number of sub-areas in the facula regions actually gives a roughly linear relationship between the oscillation amplitude and field strength. From Fig.  5c, we can conclude following empirical formula for the oscillating amplitude in the sufficiently magnetized regions ($B_0 \ge 10$ G) outside sunspots 
\begin{equation}
B_{1z} = 0.35 + c (B_{0} - 10) \, G
 \label{v12Z1}
\end{equation} 
where c (= $1.8 \times 10^{-3}$) is the linear regression coefficient that should be related to the constant $\sigma$. For the LOS magnetic field less than 10 G, the oscillation amplitude gradually falls into a constant noise level which is given as roughly 0.35 G in this paper (Fig.  5).  It is worth noting that the coefficient is quite small. Thus, for the plage region with 200 G magnetic field, the empirical formula gives the RMS amplitude as 0.8 G, which is usually {taken as noises}.  

\begin{figure}
\includegraphics[width=11cm]{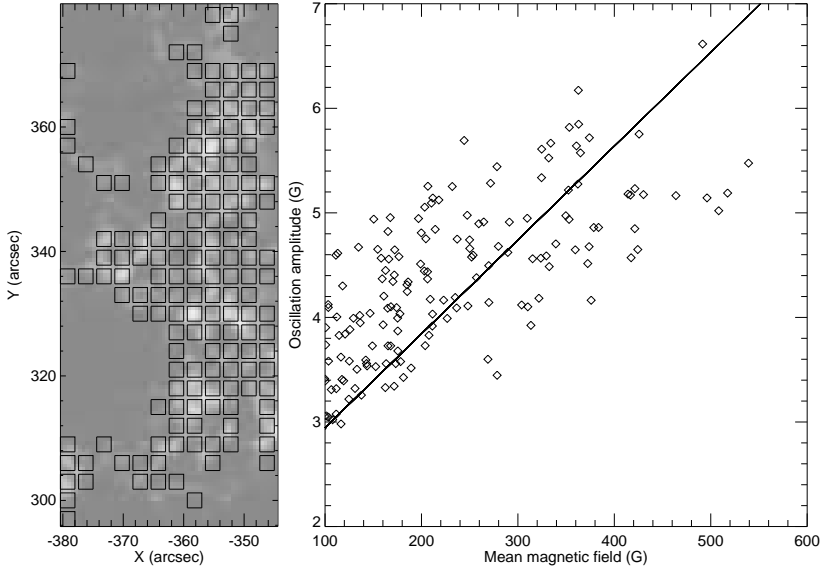}
\caption{Pixels with the stronger magnetic field are isolated in the plage region to confirm the positive relationship between the amplitude of magnetic field perturbations and field strength. Left panel shows all selected small regions ($5 \times 5$ pixels$^2$) with timely averaged magnetic field strength larger than 100 G. The field of view of the left panel is shown in Fig. 2 with a black box. The right panel gives the scatter plot for the RMS amplitude and mean field strength from all the isolated pixels in the left panel.}
\end{figure}

The coefficient depends on the size of sub-areas. If a sub-area contains both weak and strong magnetic field pixels, the averaged oscillation amplitude will be reduced and the phase will be affected (Fig. \ref{cross}) since the signals in the weak field pixels are in anti-phase with the signals that arise from the stronger field pixels. We isolate strong magnetic field pixels in the plage region. The region for isolating strong magnetic field pixels, with positive magnetic polarity, is the black boxed area in Fig. 2.  Fig. 10a shows all selected small regions ($5 \times 5$ pixels$^2$) with timely averaged magnetic field larger than 100 G.  A scatter plot for the RMS amplitude and mean magnetic strength from all selected sub-areas in Fig. 10a is given in the right panel. The plot again shows that the perturbation amplitude grows with the magnetic field, but with a larger linear regression coefficient as $9.0 \times 10^{-3}$.  

 It looks unreasonable for the positive relationship, since the stronger magnetic field seems more difficult to perturb. Nevertheless, observations show that p-mode power is substantially suppressed in magnetic regions, including sunspots which are the extreme cases due to the strongest magnetic field \citep{Lites1982, Title1992, Jain1996}. Many mechanisms have been proposed (e.g., Jain \& Haber 2002; Jain et al. 2009 and references therein). We see that, for plage regions with a high $\beta$ value, the mechanism might be simple. For small perturbations in magnetized fluid, $\vec{\nabla} p_1$ will redistribute itself to overcome the gradient of magnetic pressure, which is what equation (\ref{deviation1}) means. A small part of energy given by $\vec{\nabla} p_1$ will thus be converted into magnetic oscillations. For the linearized MHD wave equations in this paper, we can get solutions for $B_{1z}$ when we include the mass conservation equation and the energy equation. However, in order to take advantage of high-$\beta$ condition in the plage region, i.e., the condition expressed with equation (\ref{QJ12cJ}),  we have taken a different approach. In this way, the general solution given by equation (\ref{v12T}) can help us to understand the progressive behavior of amplitude of magnetic oscillations when magnetic field strength deviates slightly from zero in a high-$\beta$ condition. As we have seen, phase relationship between magnetic oscillations and Doppler velocity can also be deduced. However, we can not use equation (\ref{v12T}) for sunspots, where magnetic field dominates, to deduce the similar conclusions. From Fig. 6a and c, we see that the amplitude of oscillations in sunspots is obviously lower than that of plage regions with similar magnetic field strength.  

We have shown that He \textsc{i} 10830 \AA\ absorption in the moss region exhibits an oscillating behavior and the oscillations are associated with the magnetic perturbations. The association also supports that the observed magnetic perturbations are signals of magneto-acoustic oscillations in the moss region. He \textsc{i} 10830 \AA\ is a line from upper chromosphere, and its absorption signifies EUV emission, the association will be very useful for studying upward mass and energy flows from the photosphere. The 5-min periodic oscillations of He \textsc{i} 10830 \AA\ absorption in the moss region are apparently the result of p-mode leakage. The oscillations from p-mode leakage have been found in coronal loops \citep{Jess2015, Banerjee2007, Nakariakov2005, 1999SoPh..190..249N}.  
  
 As for the specific physical heating mechanism, the magneto-acoustic disturbances may evolve into upward propagating shocks \citep{Hansteen, 2015ApJ...799L...3R} or they may even modulate ongoing small-scale magnetic reconnection \citep{ChenPriest2006, Tian2019}.  There is a potential for p-modes to be converted to Alfv{\'e}nic waves (e.g., Morton et al. 2019), which may provide another contribution to coronal heating. Search for the signature of Alfv{\'e}n waves in the magnetic oscillations will be an important topic for current high resolution observations on ground (Cao et al. 2010; Liu et al. 2014). Also, future ground-based high-resolution spectro-polarimetry observations and space missions for the chromosphere and transition region, the interface layer, will help to determine which heating mechanism is working \citep{2019arXiv191208650S, 2012SciSc..42.1282L, 2017TIANRAA....17..110T}. Since photospheric magneto-acoustic oscillations can leak into the chromosphere and corona, the coronal waves studied in coronal seismology can be traced back to their photospheric source. In this way, coronal seismology will be connected to traditional helioseismology. This topic is also worthy of further investigations. 

\acknowledgements {We thank the anonymous referee for helping us to improve the paper and SDO/AIA, SDO/HMI teams for providing the valuable data. SDO is a NASA project. The HMI data are downloaded via the Virtual Solar Observatory and the Joint Science Operations Center. This work was supported by NSFC grants 11790302, 11729301, 11773061, 11773072 and 11873027 and the project of 2017XBQNXZ-A-007.  W. Cao acknowledge the support AFOSR-FA9550-19-1-0040 and NASA-80NSSC20K0025 grants. Prof. Qiu, Jiong from Montana State University assisted us with some of the MHD formulae. High-resolution He \textsc{i} 10830 \AA\ narrow band imaging was carried out with the 1.6-meter aperture Goode Solar Telescope at BBSO. We thank Prof. Goode for sharing the telescope's observing time with us. BBSO operation is supported by NJIT and US NSF AGS-1821294 grant. GST operation is partly supported by the Korea Astronomy and Space Science Institute, the Seoul National University, the KLSA-CAS and the Operation, Maintenance and Upgrading Fund of CAS for Astronomical Telescope and Facility Instruments.}

\end{document}